\documentclass[aps,prl,twocolumn,groupedaddress,showpacs]{revtex4-1}
\usepackage{graphicx}
\begin{document}

% Use the \preprint command to place your local institutional report
% number in the upper righthand corner of the title page in preprint mode.
% Multiple \preprint commands are allowed.
% Use the 'preprintnumbers' class option to override journal defaults
% to display numbers if necessary
%\preprint{}

%Title of paper
\title{3D projection sideband cooling}

% repeat the \author .. \affiliation  etc. as needed
% \email, \thanks, \homepage, \altaffiliation all apply to the current
% author. Explanatory text should go in the []'s, actual e-mail
% address or url should go in the {}'s for \email and \homepage.
% Please use the appropriate macro foreach each type of information

% \affiliation command applies to all authors since the last
% \affiliation command. The \affiliation command should follow the
% other information
% \affiliation can be followed by \email, \homepage, \thanks as well.
\author{Xiao Li}
%\email[]{Your e-mail address}
%\homepage[]{Your web page}
\thanks{Present address: Joint Quantum Institute, University of MD, College Park, MD 20742 USA}
\author{Theodore A. Corcovilos}
\author{Yang Wang}
\author{David S. Weiss}
\email{dsweiss@phys.psu.edu}
%\altaffiliation{}
\affiliation{Physics Department, The Pennsylvania State University, 104 Davey Laboratory, University Park, Pennsylvania 16802 USA}

%Collaboration name if desired (requires use of superscriptaddress
%option in \documentclass). \noaffiliation is required (may also be
%used with the \author command).
%\collaboration can be followed by \email, \homepage, \thanks as well.
%\collaboration{}
%\noaffiliation

\date{\today}

\begin{abstract}
We demonstrate 3D microwave projection sideband cooling of trapped, neutral atoms. The technique employs state-dependent potentials that enable microwave photons to drive vibration-number reducing transitions. The particular cooling sequence we employ uses minimal spontaneous emission, and works even for relatively weakly bound atoms. We cool 76\% of atoms to their 3D vibrational ground states in a site-resolvable 3D optical lattice.
\end{abstract}

% insert suggested PACS numbers in braces on next line
\pacs{37.10.De, 37.10.Jk}
% insert suggested keywords - APS authors don't need to do this
%\keywords{}
\maketitle

%body of paper here - Use proper section commands
% References should be done using the  \cite, \ref, and \label commands
Progress in physics often follows progress in cooling. For instance, the development of laser cooling \cite{Ref1} of atoms in the 1980s led to dramatically improved atomic clocks \cite{Ref2} and to new types of measurements and devices, like accelerometers \cite{Ref3} and gyroscopes \cite{Ref4}. The application of evaporative cooling to alkali-metal atoms \cite{Ref5,Ref6} in the early 1990s led to the creation of Bose-Einstein condensates \cite{Ref6,Ref7},  and degenerate Fermi gases \cite{Ref8}. It also relegated laser cooling to a critical but merely intermediate step in many cold atom experiments. Evaporatively cooled atoms have been used to observe a wealth of weakly coupled gas phenomena, as well as such strong coupling phenomena as superfluid-Mott insulator transitions, the BEC-BCS crossover, and 1D and 2D gases \cite{Ref9}. Some major goals, however, like implementing models of quantum magnetism \cite{Ref10,Ref11},  and high-$T_c$ superconductivity \cite{Ref12} require still better cooling, and particularly cooling that works on atoms in an optical lattice. Our 3D projection sideband cooling technique works well on atoms that are only weakly in the Lamb-Dicke limit, so it can be applied to large spacing optical lattices or other optical traps where the occupancy of individual sites can be measured, so site occupation is not a source of entropy. Conceptually similar to Raman sideband cooling \cite{Ref13,Ref14,Ref15}, our technique differs in that it uses a state-dependent potential that allows the Raman laser pulses to be replaced by microwaves, adiabatic rapid passage \cite{Ref16}, independent cooling of each spatial direction, and employment of a sequence with the least possible spontaneous emission. After 3D projection sideband cooling, 76\% of the atoms are in their absolute 3D vibrational ground state.

The absolute ground state occupation, $P_0$, for atoms in deep optical lattices has been made very close to 1 by the elegant approach of evaporative cooling to create a quantum degenerate gas, followed by adiabatic turn-on of an optical lattice past the Mott insulator transition \cite{Ref17}. Nonzero cooling temperatures and nonadiabaticity lead to site occupation defects. Observing these defects removes the site distribution entropy, but the observation involves polarization gradient laser cooling (PGC), which dramatically decreases $P_0$ \cite{Ref18,Ref19}. There have been several proposals \cite{Ref20,Ref21,Ref22}  and one experiment \cite{Ref23} to heal defects without observing them, but these have not yet been demonstrated to improve overall site occupation errors. Better laser cooling is critical to an alternative way to minimize the total entropy per particle, by observing and correcting filling defects, and then laser cooling in the lattice. The ability to cool $P_0$ close to 1 after observing defects is especially important for quantum computing experiments, since unknown occupancy defects lead to severe error  \cite{Ref24} and high temperatures lead to inhomogeneous broadening of gate transitions and high heating rates.

For laser cooling to leave atoms colder than a photon recoil energy, $E_r$, it must direct atoms toward a state that is not excited during cooling. Atoms irreversibly enter this dark state via spontaneous emission and accumulate there. There is no fundamental limit to $P_0$. For single ions in Paul traps, Raman sideband cooling routinely achieves $P_0 = 0.995$ \cite{Ref25}.  Because atoms in optical lattices are trapped much less tightly than ions in Paul traps and because the optical lattice itself can compromise the dark state in optical lattices, 3D Raman sideband cooling has previously only reached $P_0 = 0.56$ \cite{Ref15}, and that was in 400-nm scale lattices, where site occupation has not been resolved. 

Projection sideband cooling \cite{Ref30}, a version of which was recently demonstrated in 1D \cite{Ref26}, accomplishes coherent transfer to a lower vibrational level, $n$, without relying on the momentum of the transferring beam. It requires a state-dependent potential, which we create by rotating the linear polarization of one of a pair of optical lattice beams. This shifts the trap centers for atoms in different magnetic sublevels so that each vibrational wavefunction associated with one magnetic sublevel has a nonzero spatial projection on all the vibrational wavefunctions associated with the other magnetic sublevel. In the resolved sideband limit, microwave photons (or copropagating Raman beams) can therefore drive vibrational transitions directly between any two vibrational levels.

%% Fig 1 - Atom images
\begin{figure}
\includegraphics[width=3.4in,clip]{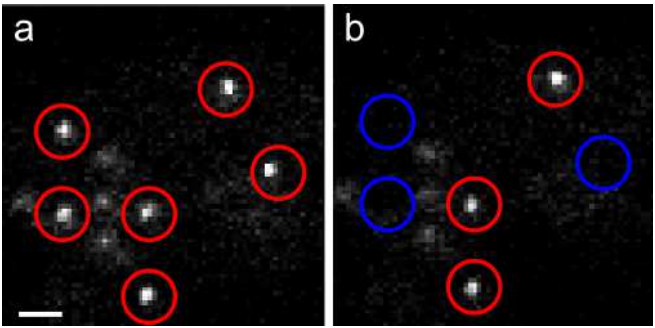}%
\caption{\label{fig1}(Color online) Detection photos.  Individual atoms (highlighted by red circles) within the image plane are detected by imaging scattered PGC light.  We typically take successive pictures of 5 planes.  We take an image of each plane (a) before cooling.  After cooling the atoms, applying a microwave AFP spectroscopy pulse, and clearing $F = 4$ atoms, we take a second image of each plane (b).  New vacancies are indicated by blue circles.  The ratio of the number of atoms in (b) compared to (a) measures the microwave transfer efficiency.  The scale bar is one lattice spacing ($4.9\;\mathrm{\mu m}$). These low-occupancy images are shown for clarity, but site occupancies over 40\% can be routinely attained.}
\end{figure}

Our apparatus is largely described in Ref.~\cite{Ref27}. We form a 3D optical lattice with lattice spacing $L = 4.9\;\mathrm{\mu m}$ using three pairs of blue-detuned 847.8-nm beams (55 mW per beam and $1/e^2$ beam radii of 65 $\mathrm{\mu m}$) that are $10^\circ$ from copropagating and linearly polarized perpendicular to their plane of incidence.  The path lengths of the two beams in each pair are matched to minimize relative phase fluctuations.  The pairs are shifted in frequency relative to each other using acousto-optic modulators to prevent interference among lattice pairs. The vibrational frequencies of the individual lattice sites, $\nu_i$, are 16, 16, and 15 kHz in the $i = x,\;y,\;\textrm{and}\;z$ directions, respectively.  Cesium atoms are loaded from a magneto-optic trap and imaged with a 0.55 numerical aperture lens using fluorescence from PGC light.  PGC keeps the atoms' temperature low enough ($\mathord{\sim} 5\;\mathrm{\mu K}$ or $\left< n_i \right> \sim 6$ in each direction) that they very rarely thermally hop over the $165\;\mathrm{\mu K}$ lattice barriers \cite{Ref27}. The imaging depth of field is short enough ($3\;\mathrm{\mu m}$) to allow measurement of the site occupancy, which is always either zero or one, of all lattice sites in a plane [see Fig.~\ref{fig1}(a)]. Translating the lens axially allows multiple planes to be successively imaged.

%% Fig. 2 - Cooling sequence cartoon
\begin{figure}
\includegraphics[width=3.4in]{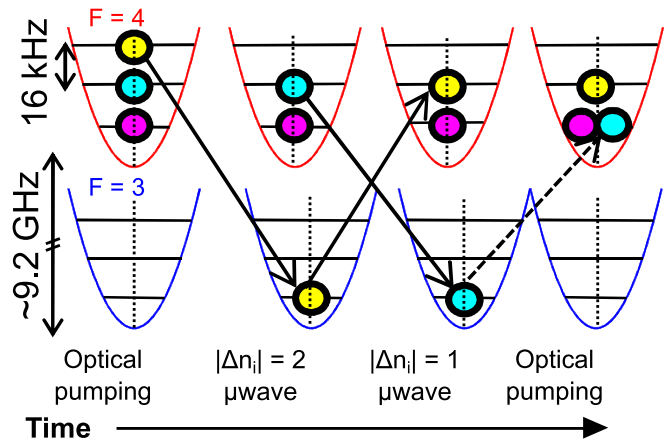}%
\caption{\label{fig2}(Color online) The steps of the projection sideband cooling sequence. The drawing shows a representation of the harmonic oscillator potential of the $F = 4$, $m_F=-4$ (upper row) and $F = 3$, $m_F=-3$ (lower row) hyperfine states.  The vertical dotted lines denote the centers of the potentials, which are displaced relative to each other during the microwave pulses.  The hyperfine levels are separated in energy by $\mathord\sim 9.2\;\mathrm{GHz}$ and the vibrational levels associated with each sublevel are separated by $\mathord\sim 16\;\mathrm{kHz}$.  The balls represent atoms
% initially in the ground and first two excited states with $F = 4$,
 using colors to help track the atoms through the sequence.  Note that in the experiment each lattice site contains at most 1 atom.  Time flows left to right, showing each step in the cooling sequence.  The solid arrows represent adiabatic microwave transitions, first $|\Delta n_i| = 2$ followed by $|\Delta n_i| = 1$.  The dashed arrow represents optical pumping from the $F = 3$ hyperfine level to the $F' = 4$ excited state (not shown) and decay back to the $F = 4$ state.  Most of the time (73\% in the ground state), optical pumping does not change that atom's vibrational level.  After the second optical pumping step, the sequence is repeated with a new lattice axis.}
\end{figure}

The projection sideband cooling sequence consists of the following steps, described in detail below: optically pump, shift the lattice, apply two microwave pulses, unshift the lattice, and repeat.  The goal is to transfer atoms to the $F = 4, m_F = -4, n_i = 0$ dark state.
We begin projection sideband cooling (see Fig.~\ref{fig2}) by optically pumping atoms into the $F = 4$, $m_F = -4$ stretched state.  A uniform magnetic field of 650 mG is applied parallel to the optical pumping beam propagation direction ($\frac{1}{\sqrt{2}}(\hat{x}+\hat{y})$) to define the quantization axis and split the Zeeman sublevels \cite{SM3}.  Next, using an axial electro-optic modulator followed by a quarter-wave plate, we tilt the linear polarization of one $i\  (= x, y,\textrm{ or }z)$ beam by $5.4^\circ$ in 1.5 ms so that the potentials for the $F = 4$, $m_F = -4$ and $F = 3$, $m_F = -3$ states are relatively displaced by 35 nm ($\sim$ half of the harmonic oscillator length).  This shift makes the spatial overlap integrals $\left| \left< n_i | n_i - 1\right>\right|$ and $\left| \left< n_i | n_i - 2\right>\right|$
%greater than $\mathord{\sim} 0.1$ for $n_i < 11$ and 20 respectively
large enough to support microwave transitions \cite{SM1}.
%, as shown in Fig.~\ref{fig3}.
We then apply an adiabatic fast passage (AFP) microwave pulse resonant with the $F = 4, n_i \leftrightarrow F = 3, n_i-2$ transition  (hereafter, $\left| \Delta n_i \right| = 2$) .  A second AFP pulse is then applied at the $F = 3, n_i \leftrightarrow F = 4, n_i + 1$ frequency (``$\left| \Delta n_i \right| = 1$ ''), after which the lattice polarization is returned to its initial angle in 1.5 ms.  Next, the sequence is repeated using a different axis of lattice translation.  As $P_0$ approaches 1, this sequential cooling is much more efficient than cooling a superposition of motional eigenstates and waiting for coherent evolution into other superpositions, especially if the coherent evolution does not provide equal mixing among states. One full cooling cycle takes $T= 30\;\mathrm{ms}$.

The microwave AFP pulses we use have power $P(t)$ and frequency $f(t)$ varying with time $t\in[0,\tau]$ as $P(t)= P_\mathrm{max} \sin^4\frac{\pi t}{\tau}$ and $f(t)=f_0+ \Delta f\;\mathrm{sgn}(\frac{t}{\tau}-\frac12 )\sqrt{1-\sin^4\frac{\pi t}{\tau}}$ where $P_\mathrm{max} = 3\;\mathrm{W}$ is the maximum power%
%(corresponding to a maximum Rabi frequency of $2\pi \times 11\;\mathrm{kHz}$)%
, $\Delta f = 4\;\mathrm{kHz}$ is the chirp range, and $\tau = 3\;\mathrm{ms}$ is the pulse length.  The AFP pulse has a measured transfer efficiency of 96(2)\%  (parentheses indicating the $1\sigma$ statistical uncertainty in the final digit) and is insensitive to inhomogeneous frequency broadening (up to half of $\Delta f$) and variations in the transition matrix element \cite{SM1}.

%%% Fig. 3 - Matrix elements plot
%\begin{figure}
%\includegraphics[width=3.4in]{fig3}%
%\caption{\label{fig3}(Color online) Overlap amplitude $\left| \left< n_i | n_i + \Delta n_i \right>\right|$ when the initial and final states have a relative spatial displacement of 0.59 of the harmonic oscillator length.  The horizontal dashed line represents the minimum value needed for $>98\%$ microwave transfer using adiabatic fast passage with our experimental parameters. Note that at least one of the $\Delta n_i = 1$ or 2 amplitudes is above the threshold for all values of $n_i$.}
%\end{figure}

As in all sideband cooling, the two-pulse sequence does not affect atoms that are initially in the dark state ($F = 4, m_F=-4,  n_i=0$). The pulses reduce $n_i$ by one for all other atoms. But unlike for a single $\left|\Delta n_i\right| = 1$ pulse, all atoms end up in $F = 4$ except for those that are initially in $n_i = 1$ and others that fail to make one of the AFP transitions. The sequence thus minimizes spontaneous emission. For instance, if the AFP pulses were perfect, atoms from high lying $n_i$ levels would need only one successful optical pumping cycle per direction to reach the dark state, and these would be from the $n_i = 0$ state, from which atoms are least likely to change vibrational states during optical pumping. By minimizing stochastic fluctuations in $n_i$, cooling is accomplished in fewer steps. This is especially important when a system is not well in the Lamb-Dicke limit, $\eta = \sqrt{E_r/h\nu} \ll 1$, where $E_r$ is the optical pumping photon recoil energy. In our experiment, where $\eta = 0.37$ and it takes an average of three spontaneous emissions to optically pump, there is a 27\% probability of $n_i$ changing due to optical pumping from $n_i=0$, a probability that increases with initial $n_i$.

\begin{figure}
\includegraphics[width=3.4in]{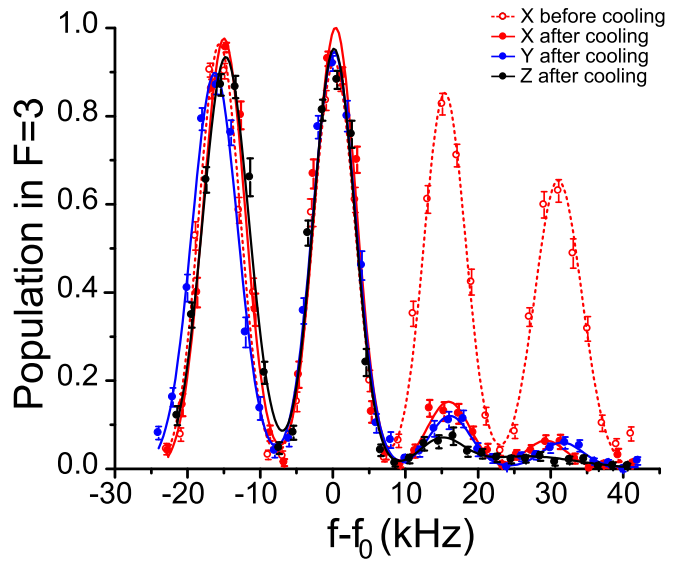}%
\caption{\label{fig4}(Color online) Microwave spectra before and after projection sideband cooling.  The carrier frequency of the transition $F = 4$, $m_F = -4, n_i$ to $F = 3$, $m_F = -3, n_i$ is $f_0=9.191215\;\mathrm{GHz}$. The experimental data is the fraction of atoms that are transferred to the $F = 3$ state by a microwave AFP pulse as a function of its center frequency, $f$ (see Fig.~\ref{fig1}).  The open red circles with dashed line are the $x$-direction spectrum before projection sideband cooling.  Before spectra of $y$ and $z$ (not shown) are similar to the one for $x$.  The filled circles with solid lines are the spectra after 35 cooling cycles in the $x$ (red), $y$ (blue) and $z$ (black) directions.  The curves are four-peak Gaussian fits to the data.  The peaks correspond, from left to right, to the $\Delta n_i = +1,\;0,\;-1,\;\textrm{and}\;\mathord-2$ vibrational transitions.  The error bars are $1\sigma$ statistical errors from averaging $\mathord\sim 300$ atoms per point.}
\end{figure}

\begin{figure}
\includegraphics[width=3.4in]{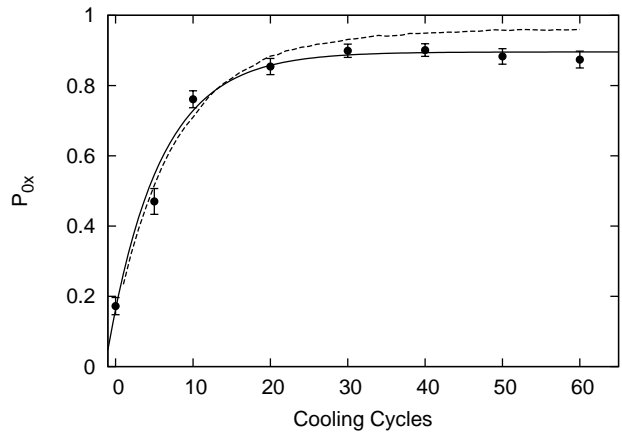}%
\caption{\label{fig5} Ground state occupancy vs number of cooling cycles.  The filled circles indicate the ground state occupancy in the $x$ lattice direction with statistical $1\sigma$ error bars.  The solid line shows an exponential fit to these data.  The dashed line shows the results of a Monte Carlo simulation of 10\,000 atoms containing the known experimental factors. The simulation yields an asymptotic ground state occupancy of 0.957(1), with a fit $1/e$ time of 8.47(4) cycles. }
\end{figure}

We measure $P_{0i}$ using microwave spectroscopy.  We rotate one lattice beam polarization, apply an AFP pulse, and then push the $F = 4$ atoms from the lattice with a $13$--$\mathrm{mW/cm^2}$ laser beam resonant with the $F = 4$ to $F' = 5$ cycling transition for $100\;\mathrm{\mu s}$. We then count the atoms that were transferred to the $F = 3$ state by the AFP pulse using a fluorescence image [Fig.~\ref{fig1}(b)].  Figure \ref{fig4} shows typical spectra obtained by scanning the center frequency of the AFP pulse, before and after projection sideband cooling. The carrier ($\Delta n_i = 0$) and sideband ($\Delta n_i \neq 0$) transitions are resolved. The fraction of atoms that remains after a $\Delta n_i = -1$ microwave pulse, normalized by the size of the $\Delta n_i = +1$ sideband, indicates the fraction that started in $n_i > 0$. Similarly, the $\Delta n_i = -2$ sideband counts atoms initially in $n_i > 1$. The steady state projection-cooled values are reached within $\mathord{\sim} 25T$ (see Fig.~\ref{fig5}). $P_{0i}$ is 0.90(2) in the $x$ and $y$ directions and 0.94(3) in the $z$ direction.  The cooling is anisotropic because the optical pumping beam is in the $x$-$y$ plane, and so disproportionately heats in those directions.  After cooling, $P_0 = P_{0x} P_{0y} P_{0z} = 0.76(3)$, which roughly corresponds to an in-lattice temperature of 300 nK (the distribution is not quite thermal).

%% Fig 4 - Microwave spectra

%% Fig 5 - Cooling v. time

The steady state $P_0$ comes from a balance of cooling and heating rates.  We measure the probability of vibrational excitation due to the lattice by using an AFP pulse and clearing sequence that leaves only ground state atoms, waiting, optically pumping back to $F = 4$, and then measuring $n_i>0$.  We find a 0.020(5) excitation probability per direction per $T$. By holding atoms in the lattice and counting the atoms that depump into the $F = 3$ state, we independently measure the optical scattering rate to be 0.08(3) per $T$, which is 13 times more than expected from our blue-detuned standing waves. We suspect the additional scattering is caused by a residual traveling wave component. The measured photon scattering rate accounts for the vast majority of the measured heating out of the dark state. Laser beam measurements show that technical noise (intensity, pointing, and lattice phase fluctuations) heats much less ($<4\times10^{-4}$, $<4\times10^{-5}$, and $<6\times10^{-3}$ per $T$ (30 ms),  respectively). We also measure the probability that cooling inadvertently excites out of the dark state by repeating isolated steps of the cooling cycle, including lattice translation, optical pumping, and AFP pulses. These cooling steps each have $<10^{-3}$ probability per $T$ of heating atoms out of the dark state. The circular polarization quality of the optical pumping pulse is the closest imperfection to being a problem; we ensure that there is $<0.1\%$ of the wrong circular polarization. 

We have performed Monte Carlo cooling calculations using the measured heating rates. Figure \ref{fig5} compares these calculations with the experimental $P_{0x}$ as a function of the number of cooling cycles. An exponential fit to the experimental data gives a $1/e$ time constant of $6.8(8)T$ and a steady state $P_{0x}$ of 0.90(1). The simulation yields 0.957(1), a discrepancy we have not understood.  Still, it is fairly clear how to improve on these cooling results. Spontaneous emission, the dominant heating source, decreases inversely with the available lattice power if the lattice depth is kept constant by increasing the detuning.  We currently use only 55 mW per beam, so an order of magnitude increased detuning is technically viable. Improved mirror damping can improve the lattice phase fluctuations four-fold, so that it does not become a problem. The probability that the final optical pumping does not succeed in putting atoms in $n = 0$ is proportional to $\eta^2 \propto L$, with a proportional decrease in $1-P_0$ \cite{SM2}.  A 3D array's $L$ can be halved (from our value of $4.9\;\mathrm{\mu m}$) and still imaged, and 2D arrays have been imaged with 9 times smaller $L$ \cite{Ref19}. Scaling from our experimental results, we infer that with these trap changes, and without including a possible improvement from shorter $T$, one could achieve $P_0 > 0.98$ and $> 0.995$ for 3D cooling in 3D and 2D site-resolvable lattices respectively. Cooling would remain in the \emph{festina lente} limit (optical pumping rate $\ll  \nu$)\cite{Ref28,Ref29}, so rescattered optical pumping light would not be a source of heating.

In conclusion, we have shown that 3D projection sideband cooling with a two-pulse AFP sequence efficiently lowers a trapped atom's vibrational energy, making it effective at cooling in weak optical traps. We have obtained ground state occupancies that are higher than have been obtained in much tighter lattices. The technique can be used in conjunction with single site imaging resolution to initialize a neutral atom quantum computer. In more closely spaced optical lattices, site occupancy determination followed by projection sideband cooling might be competitive with evaporative cooling as a way to get the lowest entropy atomic ensembles for quantum simulations.

\begin{acknowledgments}
We gratefully acknowledge funding from DARPA.
\end{acknowledgments}

% Create the reference section using BibTeX:
%\bibliography{3dpc-4}
%merlin.mbs apsrev4-1.bst 2010-07-25 4.21a (PWD, AO, DPC) hacked
%Control: key (0)
%Control: author (8) initials jnrlst
%Control: editor formatted (1) identically to author
%Control: production of article title (-1) disabled
%Control: page (0) single
%Control: year (1) truncated
%Control: production of eprint (0) enabled
%

\end{document}